\documentclass[twocolumn,times,tighten,trackchanges]{aastex62}

\usepackage{natbib}
\usepackage{xspace}
\usepackage{amsmath}
\usepackage{url}
\newcommand{\HII}{\mbox{H\,{\sc ii}}\xspace}
\newcommand{\NaID}{\mbox{Na\,{\sc i}\,D}\xspace}

\newcommand{\MgII}{\mbox{Mg\,{\sc ii}}\xspace}

\newcommand{\SiII}{\mbox{Si\,{\sc ii}}\xspace}

\newcommand{\SiIV}{\mbox{Si\,{\sc iv}}\xspace}
\newcommand{\CIIx}{\mbox{C\,{\sc ii}}}
\newcommand{\CII}{\mbox{C\,{\sc ii}}\xspace}
\newcommand{\CIIIx}{\mbox{C\,{\sc iii}}}

\newcommand{\OI}{\mbox{O\,{\sc i}}\xspace}
\newcommand{\OIIx}{\mbox{O\,{\sc ii}}}

\newcommand{\OIIIx}{\mbox{O\,{\sc iii}}}

\newcommand{\Ha}{\mbox{H$\alpha$}\xspace}

\newcommand{\Lya}{\mbox{Ly$\alpha$}\xspace}

\newcommand{\SFR}{\ensuremath{\mathrm{SFR}}}

\newcommand{\kms}{$\mathrm{km\ s^{-1}}$}
\newcommand{\Mo}{$\mathrm{M_\sun}$}
\newcommand{\Moyr}{$\mathrm{M_\sun\ yr^{-1}}$}

\newcommand{\pix}{pixel${^{-1}}$}
\newcommand{\mum}{$\mathrm{\mu m}$}
\newcommand{\kpc}{$\mathrm{kpc}$}

\newcommand{\vmax}{\ensuremath{v_\mathrm{max}}}
\newcommand{\vout}{\ensuremath{v_\mathrm{out}}}
\newcommand{\vcir}{\ensuremath{v_\mathrm{cir}}}
\newcommand{\vmaxSiII}{\ensuremath{\vmax^\mathrm{SiII}}}
\newcommand{\vmaxCII}{\ensuremath{\vmax^\mathrm{CII}}}
\newcommand{\vmaxSiIV}{\ensuremath{\vmax^\mathrm{SiIV}}}

\newcommand{\fescc}{\ensuremath{f_\mathrm{esc}}}
\newcommand{\Cfmax}{\ensuremath{C_f^\mathrm{max}}}
\newcommand{\EWLya}{\ensuremath{\mathrm{EW}_\mathrm{Ly\alpha}}}

\newcommand{\nn}{\mbox{--}}

\shorttitle{Fast Outflows in Early Galaxies at $z=5\nn6$}
\shortauthors{Sugahara et al.}
\begin{document}

\title{Fast Outflows Identified in Early Star-Forming Galaxies at $z=5\nn6$}

%====================ApJ====================
\author{Yuma Sugahara}
\email{sugayu@icrr.u-tokyo.ac.jp}
\affil{Institute for Cosmic Ray Research, The University of Tokyo, 5-1-5 Kashiwanoha, Kashiwa, Chiba 277-8582, Japan}
\affil{Department of Physics, Graduate School of Science, The University of Tokyo, 7-3-1 Hongo, Bunkyo, Tokyo, 113-0033, Japan}
\author{Masami Ouchi}
\affil{Institute for Cosmic Ray Research, The University of Tokyo, 5-1-5 Kashiwanoha, Kashiwa, Chiba 277-8582, Japan}
\affil{Kavli Institute for the Physics and Mathematics of the Universe (WPI), University of Tokyo, Kashiwa 277-8583, Japan}
\author{Yuichi Harikane}
\affil{Institute for Cosmic Ray Research, The University of Tokyo, 5-1-5 Kashiwanoha, Kashiwa, Chiba 277-8582, Japan}
\affil{Department of Physics, Graduate School of Science, The University of Tokyo, 7-3-1 Hongo, Bunkyo, Tokyo, 113-0033, Japan}
\affil{National Astronomical Observatory of Japan, 2-21-1, Osawa, Mitaka, Tokyo 181-8588, Japan}
\author{Nicolas Bouch\'e}
\affil{Univ Lyon, Univ Lyon1, Ens de Lyon, CNRS, Centre de Recherche Astrophysique de Lyon UMR5574, F-69230 Saint-Genis-Laval, France}
\author{Peter D. Mitchell}
\affil{Leiden Observatory, Leiden University, PO Box 9513, NL-2300 RA Leiden, the Netherlands}
\author{J\'er\'emy Blaizot}
\affil{Univ Lyon, Univ Lyon1, Ens de Lyon, CNRS, Centre de Recherche Astrophysique de Lyon UMR5574, F-69230 Saint-Genis-Laval, France}

\begin{abstract}
  We present velocities of galactic outflows in seven star-forming galaxies at $z=5\nn6$ with stellar masses of $M_* \sim10^{10.1}$ \Mo. Although it is challenging to observationally determine the outflow velocities, we overcome this by using ALMA [\CIIx]158 $\mu$m emission lines for systemic velocities and deep Keck spectra with metal absorption lines for velocity profiles available to date. We construct a composite Keck spectrum of the galaxies at $z=5\nn6$ with the [\CIIx]-systemic velocities, and fit outflow-line profiles to the \SiII$\lambda1260$, \CII$\lambda1335$, and \SiIV$\lambda\lambda1394,1403$ absorption lines in the composite spectrum. We measure the maximum (90\%) and central outflow velocities to be $\vmax=700^{+180}_{-110}$ \kms\ and $\vout= 400^{+100}_{-150}$ \kms\ on average, respectively, showing no significant differences between the outflow velocities derived with the low to high-ionization absorption lines. For $M_* \sim10^{10.1}$ \Mo, we find that the \vmax\ value of our $z=5\nn6$ galaxies is 3 times higher than those of $z\sim0$ galaxies and comparable to $z\sim2$ galaxies. Estimating the halo circular velocity \vcir\ from the stellar masses and the abundance matching results, we investigate a \vmax--\vcir\ relation. Interestingly, \vmax\ for galaxies with $M_*=10^{10.0\nn10.8}$ \Mo\ shows a clear positive correlation with \vcir\ and/or the galaxy star formation rate over $z=0\nn6$ with a small scatter of $\simeq \pm 0.1$ dex, which is in good agreement with theoretical predictions. This positive correlation suggests that the outflow velocity is physically related to the halo circular velocity, and that the redshift evolution of \vmax\ at fixed $M_*$ is explained by the increase in \vcir\ toward high redshift.

\end{abstract}


\section{INTRODUCTION}
\label{sec:intro}

The energy and momentum inputs from supernovae (SNe) and active galactic nuclei (AGNs) accelerate the inter-stellar medium (ISM) outwards, and launch galactic-scale outflows. The outflows are composed of the various ISM phases from cold molecular to hot gas \citep[e.g.,][]{Veilleux:2005}. AGN-driven outflows are thought to play an important role to quench the star-forming activity in massive galaxies, and recent observational work reports the evidence that the AGN feedback may also operate in low-mass galaxies \citep{Penny.S:2018a,Manzano-King.C:2019a}. On the other hand, SN-driven outflows are thought to affect galaxies primarily in low-mass regime. The mass, momentum, energy, and metal budgets of the outflows leaked from the star-forming galaxies are theoretically important for regulating the star-forming activity in low-mass galaxies, creating the mass-metalicity relation of galaxies, and polluting the circum-galactic medium and intergalactic medium (IGM) \citep[for a review, see][]{Somerville:2015,Dayal.P:2018a}. Thus, the outflows in star-forming galaxies driven by SNe have a large impact on the galaxy and IGM evolution. Many theoretical studies contribute to revealing the details of the outflow properties using cosmological (zoom-in) simulations \citep{Oppenheimer.B:2010a,Muratov:2015,Christensen:2016a,Ceverino.D:2016b,Mitchell.P:2018a,Nelson.D:2019a}.

In the rest-frame far-ultraviolet (FUV; 1000--2000\AA) to optical bands, metal absorption and emission lines are useful to trace the kinematics of the cold and warm outflowing gas. The outflow velocity along the line of sight is estimated with the ``down-the-barrel'' technique, which measures blueshifts of the absorption lines in the galaxy spectra \citep[e.g.,][]{Heckman:2000, Martin:2005, Rupke:2005a, Rupke:2005b, Steidel:2010, Martin:2012,Zhu.G:2015a,Heckman:2015, Chisholm:2015a, Chisholm:2016b, Chisholm.J:2017a,Roberts-Borsani.G:2019a,Concas.A:2019a}, while the outflowing gas far from the galaxy is detected with the absorption lines in the background-quasar spectra \citep[e.g.,][]{Bouche:2012, Kacprzak.G:2015a, Muzahid:2015b, Schroetter:2015, Schroetter:2016a}. Broad components in emission lines also provide a signature of outflows \citep{Cicone.C:2016a,Finley.H:2017a,Concas.A:2017a,Freeman.K:2017a}, which have also recently been observed with integral-field-units spectroscopy \citep{Davies.R:2019a,Forster-Schreiber.N:2019a,Swinbank.M:2019a}.

The outflows are ubiquitously observed in the star-forming galaxies at $z<1.5$ \citep{Weiner:2009,Chen:2010,Rubin:2014}. Their outflow velocities are probed to have a positive correlation with the star formation rate (\SFR), the stellar mass ($M_*$), and the \SFR\ surface density ($\Sigma_\SFR$) \citep[e.g.,][]{Rubin:2014,Heckman:2016,Chisholm:2016a}. \citet{Sugahara.Y:2017a} use archival spectra to show that maximum outflow velocity increases from $z\sim0$ to 2 in star-forming galaxies that are in a similar $M_*$ and \SFR\ range.

%\tabletypesize{\scriptsize} %\footnotesize{ %
\begin{deluxetable*}{ccccccccccc}
\tablewidth{0pt}
\tablecaption{Galaxy properties of seven LBGs\label{tb:1}}
\centering
\tablecolumns{11}
\tablehead{
  \colhead{\hspace{0cm}name}\hspace{0cm} \vspace{-0.2cm} &
  \colhead{\hspace{0cm}R.A.}\hspace{0cm} &
  \colhead{\hspace{0cm}Decl.}\hspace{0cm} &
  \colhead{S/N of DEIMOS spectra} &
  \colhead{\hspace{0cm}$z_\mathrm{sys}$\tablenotemark{a}}\hspace{0cm} &
  \colhead{\hspace{0cm}$z_\mathrm{sys}$ error}\hspace{0cm} &
  \colhead{\hspace{0cm}${\log(M_*/M_\sun)}$}\hspace{0cm} &
  \colhead{\hspace{0cm}${\log(SFR)}$}\hspace{0cm} &
  \colhead{\hspace{0cm}${\log(M_h/M_\sun)}$}\hspace{0cm} &
  \colhead{\hspace{0cm}${\log(\vcir)}$}\hspace{0cm} \\
\colhead{} & \colhead{} & \colhead{} & \colhead{(\pix)} & \colhead{} & \colhead{(\kms)} & \colhead{} & \colhead{(\Moyr)} & \colhead{} & \colhead{(\kms)}
}
%\decimalcolnumbers
\startdata
HZ1  & 09:59:53.25 & 02:07:05.43 & 0.348115 & 5.6885 & 9  & $10.47\pm0.13$ & $1.38^{+0.10}_{-0.06}$ & $12.9^{+0.5}_{-0.5}$ & $2.8^{+0.2}_{-0.2}$ \\
HZ2  & 10:02:04.10 & 01:55:44.05 & 0.455985 & 5.6697 & 30 & $10.23\pm0.15$ & $1.40^{+0.08}_{-0.04}$ & $12.0^{+0.5}_{-0.3}$ & $2.49^{+0.2}_{-0.08}$ \\
HZ4  & 09:58:28.52 & 02:03:06.74 & 0.616747 & 5.5440 & 9  & $9.67\pm0.21$  & $1.71^{+0.31}_{-0.19}$ & $11.4^{+0.2}_{-0.1}$ & $2.29^{+0.05}_{-0.04}$ \\
HZ6  & 10:00:21.50 & 02:35:11.08 & 0.776667 & 5.2928 & 5  & $10.17\pm0.15$ & $1.69^{+0.28}_{-0.12}$ & $11.9^{+0.4}_{-0.2}$ & $2.45^{+0.1}_{-0.07}$ \\
HZ7  & 09:59:30.48 & 02:08:02.81 & 0.275164 & 5.2532 & 20 & $9.86\pm0.21$  & $1.32^{+0.09}_{-0.04}$ & $11.6^{+0.2}_{-0.1}$ & $2.34^{+0.07}_{-0.05}$ \\
HZ8  & 10:00:04.06 & 02:37:35.81 & 0.216720 & 5.1533 & 10 & $9.77\pm0.15$  & $1.26^{+0.11}_{-0.05}$ & $11.5^{+0.1}_{-0.1}$ & $2.31^{+0.04}_{-0.03}$ \\
HZ10 & 10:00:59.30 & 01:33:19.53 & 0.625486 & 5.6566 & 9  & $10.39\pm0.17$ & $2.23^{+0.08}_{-0.08}$ & $12.5^{+0.7}_{-0.5}$ & $2.7^{+0.2}_{-0.2}$ \\
\hline
composite &   -    &     -       & 1.4      &   -    & -  & $10.08$        & $1.53$ & $11.8$ & $2.4$  \\
\enddata
\tablecomments{The raw DEIMOS spectra are downloaded from KOA (PI: P. Capak). The $z_\mathrm{sys}$, $M_*$, and \SFR\ values are drawn from \citet{Capak:2015}. The $M_\mathrm{h}$ and \vcir\ values are estimated in Section \ref{sec:sample}. The physical parameters of the composite spectrum are the truncated mean discarding the maximum and minimum values.}
\tablenotetext{a}{The systemic redshift $z_\mathrm{sys}$ is determined by the [\CIIx] 158 \mum\ emission line taken by ALMA.}
\vspace{-0.7cm}
\end{deluxetable*}

The ``down-the-barrel'' technique is also appropriate for outflow studies at $z > 2$. Unlike the emission from the outflows whose detection becomes difficult toward high redshift, the absorption can be detected with a bright background continuum source. \citet{Shapley:2003} construct composites of almost 1000 Lyman-break galaxy (LBG) spectra at $z\sim3$ to discuss the relation between the FUV spectral features and the outflow properties. Recently, \citet{Du.X:2018a} report no evolution of central outflow velocities at $z\sim2\nn4$ using composites of the rest-frame FUV spectra presented in \citet{Steidel:2003,Steidel:2004}, \citet{Reddy.N:2008a}, and \citet{Jones:2012a}. Although the \Lya profile provides us the information on the neutral-gas kinematics around Lyman alpha emitters at high redshift \citep[e.g.,][]{Erb:2014a,Shibuya:2014,Hashimoto.T:2015a,Trainor:2015a,Karman.W:2017a}, even at $z\sim6$ \citep{Ajiki.M:2002a}, it is difficult to directly estimate the outflow properties only from the \Lya profile due to its strong resonance scattering.

One of the keys to estimate outflow properties is to determine the systemic redshifts of the galaxies. At the low redshift, the systemic redshifts are measured by nebular emission lines (e.g., \Ha, [\OIIIx], and [\OIIx]), but observations of the emission lines become expensive at high redshift. Some outflow studies at $z>1.5$ conduct additional near-infrared (IR) observations \citep{Steidel:2010, Shibuya:2014}, while others determine the redshifts from \Lya emission or interstellar absorption, which includes the uncertainties based on the outflows \citep{Shapley:2003,Du.X:2018a}. Moreover, precise measurements of the systemic redshifts are challenging at $z>5$, where the strong optical emission lines fall into the mid-IR bands. Although there are several nebular emission lines in the rest-frame FUV band such as \OIIIx]$\lambda\lambda1660,1666$ and \CIIIx]$\lambda\lambda1906,1908$, these lines are weak to be detected in typical star-forming galaxies at high redshift.
This problem makes it difficult to extend the outflow studies to $z > 5$.

A solution in this paper is observations with the Atacama Large Millimeter/submillimeter Array (ALMA). Recent ALMA observations detect [\CIIx] $158$ \mum\ and [\OIIIx] $88$ \mum\ emission lines in high-$z$ galaxies \citep[e.g.,][]{Capak:2015,Inoue.A:2016a,Hashimoto.T:2019a}, which enables us to measure the systemic redshifts of the galaxies. Combining the redshift determined from the ALMA observations with deep observed-frame optical spectra, we can address the outflow properties at $z>5$.
As a case study, \citet{Pavesi.R:2016a} discuss the rest-frame FUV absorption lines in HZ10, a IR-luminous LBG at $z\simeq 5.6$, and find the blueshifts with respect to the [\CIIx] emission line.

This paper presents estimates of outflow velocities in star-forming galaxies at $z=5\nn6$ and discuss the redshift evolution of the outflows from $z\sim0$ to $6$.
Section \ref{sec:sample} describes the sample of galaxies at $z=5\nn6$. 
Section \ref{sec:analysis} explains the analysis of the absorption lines in the observed-frame optical spectra. We obtain a composite spectrum of the galaxies to measure an outflow velocity.
Section \ref{sec:results} shows the results on the outflow velocity and its redshift evolution.
Section \ref{sec:discussion} discusses relations between the outflow and galaxy properties. 
Section \ref{sec:conclusion} summarizes our conclusion.
The $\Lambda$CDM cosmology is used throughout this paper: 
$\Omega_\mathrm{M} = 0.27$, $\Omega_\mathrm{\Lambda} = 0.73$, 
$h = H_0/(100\ \mathrm{km\ s^{-1}\ Mpc^{-1}}) = 0.70$,
$n_s = 0.95$, and $\sigma_8 = 0.82$.
All transitions are referred to by their wavelengths in vacuum.

\begin{figure*}[t]
  \epsscale{1.15}
  \plotone{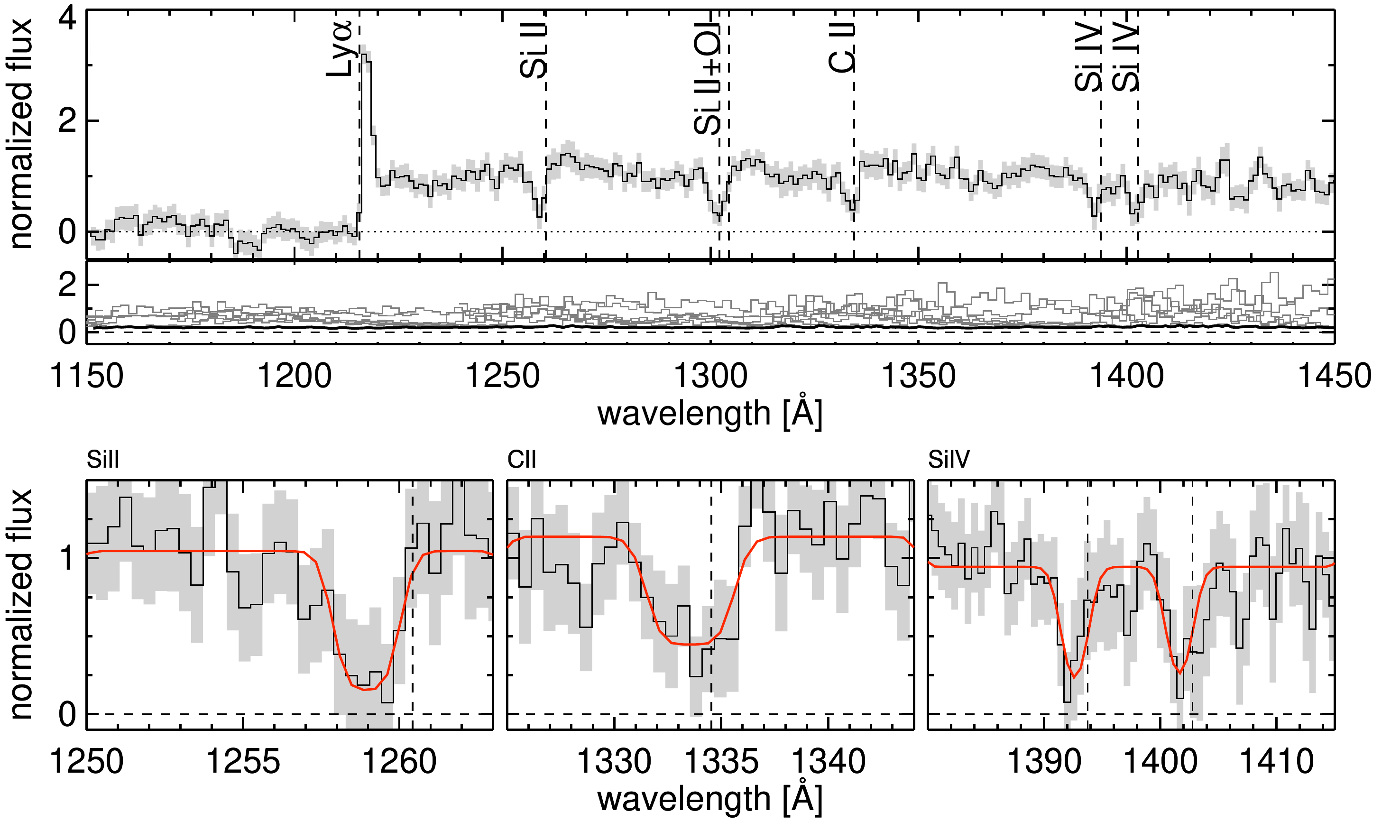}
  \caption{
    Top: composite spectrum of our sample. The gray shade indicates the 1$\sigma$ error at each pixel. The spectral resolution is smoothed for the display purpose. The rest wavelengths of the emission and absorption lines are plotted with the dashed vertical lines. The second panel under the main panel shows the 1$\sigma$ error spectra. The gray and black lines denote the errors of the normalized individual spectra and the composite spectrum, respectively. The wavelengths of the individual spectra are corrected to the rest frame using the systemic redshift determined by ALMA [\CIIx] observation. 
    Bottom: \SiII, \CII, and \SiIV absorption lines from left to right. The red solid lines are the best-fit absorption model. The vertical and horizontal dashed lines denote the rest wavelengths of the absorption lines and zero flux, respectively.
%\Lya$\lambda1216$, \SiII$\lambda1260$, \OI$\lambda1302$+\SiII$\lambda1304$, \CII$\lambda1335$, and \SiIV$\lambda\lambda1394,1403$
  }
  \label{fig:spec}
\end{figure*}

\section{Sample and Data Reduction}
\label{sec:sample}

Our sample consists of seven galaxies at $z=5\nn6$ whose spectra are taken in the optical and millimeter wavelengths. We use the galaxies presented in \citet{Capak:2015}, who observe nine LBGs and one low-luminosity quasar at $z\sim5$--$6$ in the Cosmic Evolution Survey \citep[COSMOS;][]{Scoville.N:2007a} field.
\citet{Capak:2015} obtain the rest-frame FUV spectra of the galaxies with the DEep Imaging Multi-Object Spectrograph \citep[DEIMOS;][]{Faber:2003} at the Keck II telescope. The spectroscopic configuration is the 830 \mbox{lines mm$^{-1}$} grating with the OG550 filter, which gives the wavelength coverage of 6000--9500 \AA\ and the spectral resolution of $R\sim2500\nn3500$. The total integration time is $\sim$3.5 hr for each object.

We download raw DEIMOS data of the galaxies from the Keck Observatory Archive\footnote{KOA website: \url{http://www2.keck.hawaii.edu/koa/public/koa.php.}} (KOA). The raw data are reduced with the {\tt IDL} package, the DEIMOS {\tt spec2d} pipeline, developed by the Deep Extragalactic Evolutionary Probe 2 (DEEP2) Redshift Survey team \citep{Cooper:2012,Newman:2013}. From the reduced two-dimensional multi-object-slit data, the pipeline extracts the one-dimensional spectra of the science targets. Finally, we obtain the rest-frame FUV spectra of seven out of the nine LBGs in \citet{Capak:2015}, other than two objects (HZ3 and HZ9) whose spectra we could not identify from the archive.

The ALMA follow-up observations are conducted in a project of \#2012.1.00523.S (PI: P. Capak). The Band 7 observations have detected the [\CIIx] $158$ \mum\ emission lines in all of the nine LBGs. Previous studies report possible present and past outflow signatures in the [\CIIx] emission lines of these galaxies \citep{Gallerani.S:2018a,Fujimoto.S:2019a}. In this study, we use the systemic redshifts measured from the [\CIIx] emission lines by \citet{Capak:2015}. In contrast to [\OIIIx] $88$ \mum\ or optical nebular emission lines that come from only \HII regions, the [\CIIx] emission arises from \HII regions and photo-dissociated regions, which may results in some uncertainties in the measured redshifts. However, by detecting both [\CIIx] and [\OIIIx] emission lines in objects at $z > 6$ with ALMA, recent studies reveal that redshifts determined by [\CIIx] and [\OIIIx] are consistent within the errors \citep{Marrone.D:2018a,Decarli.R:2017a,Walter.F:2018a} or show offsets less than $50$ \kms\ at most \citep{Hashimoto.T:2019a}. There is good evidence therefore that measurements of the [\CIIx] emission lines provide reliable systemic redshifts. The median redshift error is $\sim 2\times10^{-4}$, corresponding to $\sim 10$ \kms. The systemic redshifts of our galaxies are listed in Table \ref{tb:1}.

We use \SFR\ and $M_*$ derived by \citet{Capak:2015}.
The \SFR\ is estimated from the sum of the rest-frame UV and IR luminosity. The stellar mass $M_*$ is estimated from the spectral energy distribution fitting to the optical to IR photometry taken from the COSMOS photometric redshift catalog \citep{Ilbert.O:2013a} and the Spitzer-Large Area Survey with Hyper-Suprime-Cam \citep[SPLASH;][]{Steinhardt.C:2014a}. 
The halo circular velocity \vcir\ is estimated from $M_*$.
First, we convert $M_*$ into the halo mass $M_\mathrm{h}$ with the stellar-to-halo mass ratio (SHMR) given by \citet{Behroozi:2013}, who derive the SHMR at $z=0\nn8$ with the abundance matching method, although it should be noted that observational constraints are less complete and potentially less robust at $z > 4$ than at $z \lesssim 2$. Then, \vcir\ are calculated by equations in \citet{Mo:2002} expressed as
\begin{eqnarray}
  \label{eq:0a}
  \vcir &=& \left( \frac{GM_\mathrm{h}}{r_\mathrm{h}} \right)^{1/2}, \\
  \label{eq:0b}
  r_\mathrm{h} &=& \left( \frac{GM_\mathrm{h}}{100 \Omega_\mathrm{M} H_0^2} \right)^{1/3} (1+z)^{-1} , 
\end{eqnarray}
where $G$ is the gravitational constant and $r_\mathrm{h}$ the halo radius.

\section{Analysis and Measurements}
\label{sec:analysis}

Because the outflowing gas gives rise to the blueshifted metal absorption lines due to the Doppler shift, the blueshift reflects the line-of-sight velocity of the outflowing gas. The absorption-line analysis requires high signal-to-noise ratios (S/N) of the continuum spectra. Our rest-frame FUV spectra have the average S/N of $\simeq 0.47$ \pix, which is not enough for the absorption-line analysis. Therefore, we obtain a high-S/N composite spectrum by stacking all of the spectra with an inverse-variance weighted mean. The top panel of Figure \ref{fig:spec} shows the composite spectrum and its error spectrum. The continuum S/N of the composite spectrum is $1.4$ \pix\ around the \SiII$\lambda1260$ absorption line. The physical parameters of the composite spectrum are the truncated mean discarding the maximum and minimum values. We note that HZ10 has much higher \SFR\ than other galaxies. By constructing another composite spectrum without HZ10, we check whether this high-\SFR\ galaxy affects our results to confirm that our conclusion does not change.

In the wavelength range from 1150 to 1450 \AA\ in the rest frame, we use the absorption lines of \SiII$\lambda1260$, \CII$\lambda1335$, and \SiIV$\lambda\lambda1394,1403$ for the analysis, without the \SiII$\lambda1304$ line that has a nearby strong \OI$\lambda1302$ absorption line. We hereafter refer to \SiII$\lambda1260$ as \SiII.

We measure outflow velocities by fitting a line profile to the absorption lines. As the line profile, we adopt a physical profile based on the assumption of the curve of growth \citep{Rupke:2005a}. This line profile $I(\lambda)$, as a function of the wavelength $\lambda$, is expressed by
\begin{eqnarray}
  \label{eq:1}
  I(\lambda)/I_0 &=& 1-C_f+C_f\exp(-\tau(\lambda)), \\
  \label{eq:2}
  \tau(\lambda) &=& \tau_0\exp(-(v(\lambda)-v_0)^2/b^2),
\end{eqnarray}
where $I_0$ is the continuum level, $C_f$ the covering fraction, $\tau(\lambda)$ the optical depth, $\tau_0$ the optical depth at the line center, $v(\lambda)$ the velocity measured from the rest wavelength, $v_0$ the velocity at the line center, and $b$ the Doppler width. The line profile is convolved with a Gaussian profile representing the spectral resolution. The free parameters are five: $I_0$, $v_0$, $C_f$, $\tau_0$, and $b_D$. Because the composite spectrum has large noises, we treat $I_0$ as a free parameter instead of normalizing the spectrum by a stellar continuum. We fit the line profile to \SiII, \CII, and \SiIV, using an {\tt IDL} procedure {\tt MPFIT}, which performs non-linear least-squares fitting in a robust manner \citep{Markwardt:2009}. The bottom panel of Figure \ref{fig:spec} shows the best-fit model of the \SiII, \CII, and \SiIV absorption lines with the red lines.

The best-fit $v_0$ values, listed in Table \ref{tb:2}, are all significantly negative, implying that the absorption lines are blueshifted by the outflowing gas. The errors of $v_0$ are evaluated by the parametric bootstrap method. We obtain a $v_0$ distribution by fitting the line profile to 1000 resampled fluxes based on the spectral noise and use the $\pm 34$th $v_0$ values for its error. These velocities are consistent with the values estimated in the literature that analyzes the data of the same galaxies \citep{Pavesi.R:2016a,Gallerani.S:2018a}. HZ10 has the \SiII, \SiII$\lambda1304$/\OI, and \SiIV absorption lines blueshifted by $100\pm180$ \kms\ with respect to the [\CIIx] emission line \citep{Pavesi.R:2016a}. The composite emission of the [\CIIx] line in HZ1--9, without HZ5, is reported to have the broad wings that are likely generated by the outflows with the velocity of $\sigma=220\nn500$ \kms\ \citep{Gallerani.S:2018a}.

We define the maximum outflow velocity $\vmax$, the highest velocity of the outflowing gas, as 
\begin{equation}
  \label{eq:3}
  \vmax = - v_0 + b
  \sqrt{-\ln\left(\frac{1}{\tau_0}\ln\frac{1}{0.9}\right)}, 
\end{equation}
which represents the velocity where the best-fit model has a 90\% flux from the continuum to the bottome of the line.\footnote{Throughout this paper, \vmax\ is not the maximum circular velocity in the rotation curve of a galaxy or halo, which is often used in theoretical papers.}. The errors of \vmax\ are evaluated by the parametric bootstrap method that is the same as used to evaluate the $v_0$ error.

The derived maximum outflow velocities for \SiII, \CII, and \SiIV are 
$\vmaxSiII = 690^{+260}_{-120}$ \kms, 
$\vmaxCII = 720^{+140}_{-460}$ \kms, and 
$\vmaxSiIV = 610^{+240}_{-96}$ \kms, respectively. 
Low-ionized elements (\SiII and \CII) have ionization potentials lower than that of hydrogen (13.6 eV), while high-ionized elements (\SiIV) have a much higher ionization potential. Although the low- and high-ionized elements trace the different state of the ISM, $\vmaxSiII$ and $\vmaxCII$ are consistent with $\vmaxSiIV$ within the 1$\sigma$ errors. This consistency agrees with previous work on outflows at $z\sim0$ \citep{Chisholm:2016a}. 

%\tabletypesize{\scriptsize} %\footnotesize{ %
\begin{deluxetable}{cccc}
\tablecolumns{4}
\tablewidth{\columnwidth}
\tablecaption{Measured outflow velocities for the absorption lines \label{tb:2}}
\centering
\tablehead{
\colhead{redshift} \vspace{-0.2cm} & \colhead{line} & \colhead{$v_0$} & \colhead{\vmax} \\
\colhead{} & \colhead{} & \colhead{\hspace{0.2cm}(\kms)}\hspace{0.2cm} & \colhead{\hspace{0.2cm}(\kms)}\hspace{0.2cm}}
%\decimalcolnumbers
\startdata
$z=5\nn6$ & \SiII$\lambda1260$         & $  -366^{+63}_{-99} $ & $690^{+260}_{-120} $ \\
\ldots & \CII$\lambda1335$             & $  -210^{+120}_{-74} $ & $720^{+140}_{-460} $  \\
\ldots & \SiIV$\lambda\lambda1394,1403$ & $  -220^{+150}_{-100}$ & $610^{+240}_{-96} $ \\
\ldots & \SiII \& \CII                  & - & $700^{+180}_{-110} $ \\
$z\sim2$\tablenotemark{a} & \CII$\lambda1335$ & $-175^{+13}_{-11}$ & $673^{+35}_{-33}$
\enddata
%\tablecomments{}
\tablenotetext{a}{We obtain the velocities at $z\sim2$ by re-analyzing the composite spectrum of \citet{Sugahara.Y:2017a}.}
\vspace{-0.7cm}
\end{deluxetable}

\SiII and \CII have similar ionization potentials and oscillator strengths, and exhibit similar maximum outflow velocities. To obtain a typical \vmax\ value of the $z=5\nn6$ galaxies, we additionally measure the maximum outflow velocity by a simultaneous fitting to \SiII and \CII, adopting \vmax\ as a free parameter instead of $v_0$. Both lines are assumed to have the same $C_f$. The measured value is $\vmax = 700^{+180}_{-110}$ \kms. This value is consistent with $\vmaxSiII$ and $\vmaxCII$, but its error is smaller than those of \vmaxSiII\ and \vmaxCII. Table \ref{tb:2} lists the measurements of \vmax\ and $v_0$ for each absorption lines. These \vmax\ values are consistent with the results of the [\CIIx] emission analysis by \citet{Gallerani.S:2018a}. They stack the ALMA data of HZ1--9, without HZ5, to show the highest velocity of $\sim500\nn700$ \kms\ at which the [\CIIx] flux excess can be observed, although this broad flux excess may include emission from satellites around the central galaxies.

\subsection{Galaxies at $z\sim0\nn2$}

To investigate the redshift evolution of the outflow velocity, we mainly compare the measurements at $z=5\nn6$ with the results in \citet{Sugahara.Y:2017a}, who study outflows at $z\sim0\nn2$. Here, we briefly describe the sample and analysis in \citet{Sugahara.Y:2017a}. They obtain outflow velocities using optical spectra of the Sloan Digital Sky Survey Data Release 7 \citep[SDSS DR7;][]{Abazajian:2009} at $z\sim0$, the DEEP2 DR4 survey \citep{Newman:2013} at $z\sim1$, and the Keck/LRIS sample in \citet{Erb:2006c} at $z\sim2$. The galaxies at $z\sim0$ are the highly star-forming galaxies in which the absorption lines can be analyzed, while the galaxies at $z\sim1$ and $2$ are the star-forming main-sequence galaxies. The stellar masses of the galaxies are listed in Table \ref{tb:3}, which are similar to the galaxies at $z=5\nn6$ in this work.

\citet{Sugahara.Y:2017a} basically perform the same analysis as this study. The main difference in the method is to use a two-component absorption-line profile that consists of a systemic component fixed at the systemic velocity and a blueshifted component produced by outflowing gas. This difference of the fitting profile is negligible for the spectra at $z\sim0$ and $1$, where the absorption lines have small systemic components. The spectra at $z\sim2$, however, have large systemic components. We re-analyze the absorption lines of the normalized composite spectrum at $z\sim2$ to measure the maximum outflow velocity with the one-component absorption-line profile described in Section \ref{sec:analysis}. The new maximum outflow velocity becomes lower than the previous value, but the conclusions in \citet{Sugahara.Y:2017a} are not affected by this re-analysis.

Although the available absorption lines differ depending on redshifts, \citet{Sugahara.Y:2017a} carefully compare the outflow velocities at $z\sim0\nn2$ measured from different lines and show an increase in the outflow velocities from $z\sim0$ to $2$. Table \ref{tb:3} lists the \vmax\ values measured from the \NaID, \MgII, and \CII absorption lines for the $z\sim0$, $1$, and $2$ galaxy spectra, respectively.

%\tabletypesize{\scriptsize} %\footnotesize{ %
\begin{deluxetable}{ccccc}
\tablecolumns{5}
\tablewidth{0pt}
\tablecaption{stellar mass and outflow velocity of the galaxies in \citet{Sugahara.Y:2017a} \label{tb:3}}
\centering
\tablehead{
  \colhead{redshift} \vspace{-0.2cm} &
  \colhead{line\tablenotemark{a}} &
  \colhead{\vmax} &
  \colhead{$\log(M_*/M_\sun)$} &
  \colhead{symbol} \\
  \colhead{} &
  \colhead{} &
  \colhead{(\kms)} &
  \colhead{} &
  \colhead{in figures}
}
%\decimalcolnumbers
\startdata
0.064 & \NaID & 221 $\pm$ 18 & 10.2 & blue circle  \\
0.075 & \NaID & 261 $\pm$ 12 & 10.3 & blue circle  \\
0.085 & \NaID & 299 $\pm$ 11 & 10.4 & blue circle  \\
 0.11 & \NaID & 267 $\pm$ 11 & 10.5 & blue circle  \\
 0.13 & \NaID & 327 $\pm$ 10 & 10.6 & blue circle  \\
 0.14 & \NaID & 373 $\pm$ 20 & 10.8 & blue circle  \\
 1.4  & \MgII & 445 $\pm$  9 & 9.99 & cyan diamond \\
 1.4  & \MgII & 442 $\pm$ 27 & 10.4 & cyan diamond \\
 1.4  & \MgII & 569 $\pm$ 12 & 10.6 & cyan diamond \\
 2.2  & \CII  & $673^{+35}_{-33}$\tablenotemark{b} & 10.3 & green diamond \\
\enddata
\tablenotetext{a}{Absorption lines that are analyzed for estimations of the \vmax\ values.}
\tablenotetext{b}{The maximum outflow velocities at $z\sim2$ are re-measured in this study.}
\vspace{-0.7cm}
\end{deluxetable}

\section{Results}
\label{sec:results}

\subsection{Maximum Outflow Velocity vs. Galaxy Properties}
\label{sec:results1}

\begin{figure}[t]
  \epsscale{1.2}
  \plotone{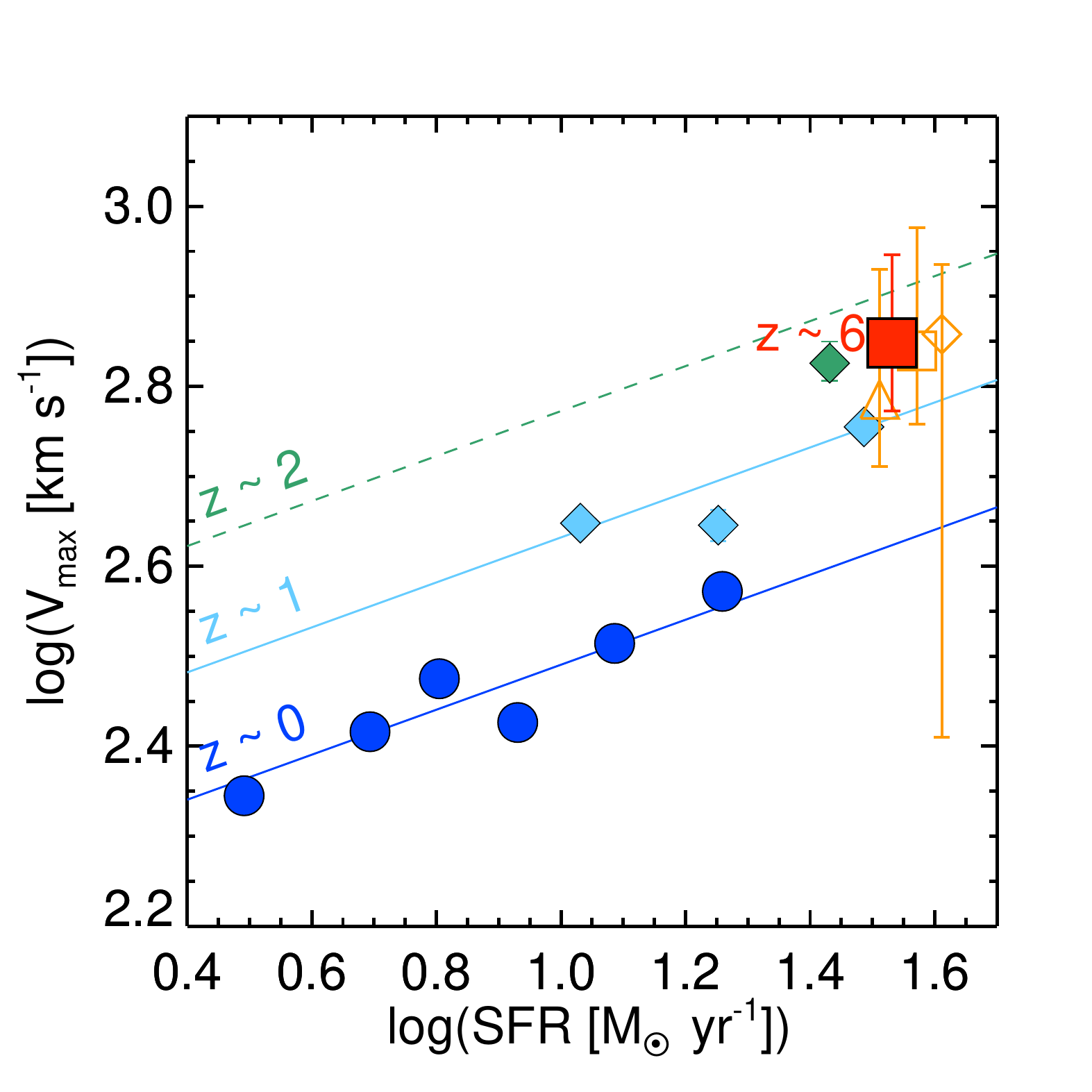}
  \caption{Maximum outflow velocity $\vmax$ as a function of \SFR\ over $z\sim0\nn6$. The filled red square indicates the $\vmax$ value at $z=5\nn6$ measured with the simultaneous fitting of the \SiII and \CII lines. The open orange square, diamond, and triangle represent the values measured for \SiII, \CII, and \SiIV, respectively. The data points at $z\sim0$ (blue) and $z\sim1$ (cyan) are presented by \citet{Sugahara.Y:2017a}. The \vmax\ value at $z\sim2$ (green) is derived from the composite spectrum in \citet{Sugahara.Y:2017a}, but re-calculated in the manner of this work. The error bars show 1$\sigma$ measurement errors. The blue, cyan, and green lines express the best-fit relations at $z\sim0$, $1$, and $2$, respectively, which have slopes fixed at the best-fit value at $z\sim0$ \citep{Sugahara.Y:2017a}. There is a offset between the green diamond and the green dashed line because the $z\sim2$ best-fit relation is the one in \citet{Sugahara.Y:2017a}.
  }
  \label{fig:res1}
\end{figure}

Figure \ref{fig:res1} shows the maximum outflow velocity as a function of \SFR. The \vmaxSiII, \vmaxCII, \vmaxSiIV, and \vmax\ values are plotted with the open orange square, diamond, triangle, and filled red square, respectively. \citet{Sugahara.Y:2017a} illustrate that the outflow velocity increases from $z\sim0$ (blue) to $2$ (green) in star-forming galaxies with similar $M_*$ and \SFR. We find that the \vmax\ value at $z=5\nn6$ is $\sim0.2$ dex higher than the relation at $z\sim0$ and comparable to the value at $z\sim2$, although the \SFR\ values at $z\sim0$ are not as high as those at $z=5\nn6$. This means that the outflow velocity shows a strong increase from $z\sim0$ to $2$ and a slight or no increase from $z\sim2$ to $6$ in galaxies with similar $M_*$ and \SFR.

\begin{figure}[t]
  \epsscale{1.2}
  \plotone{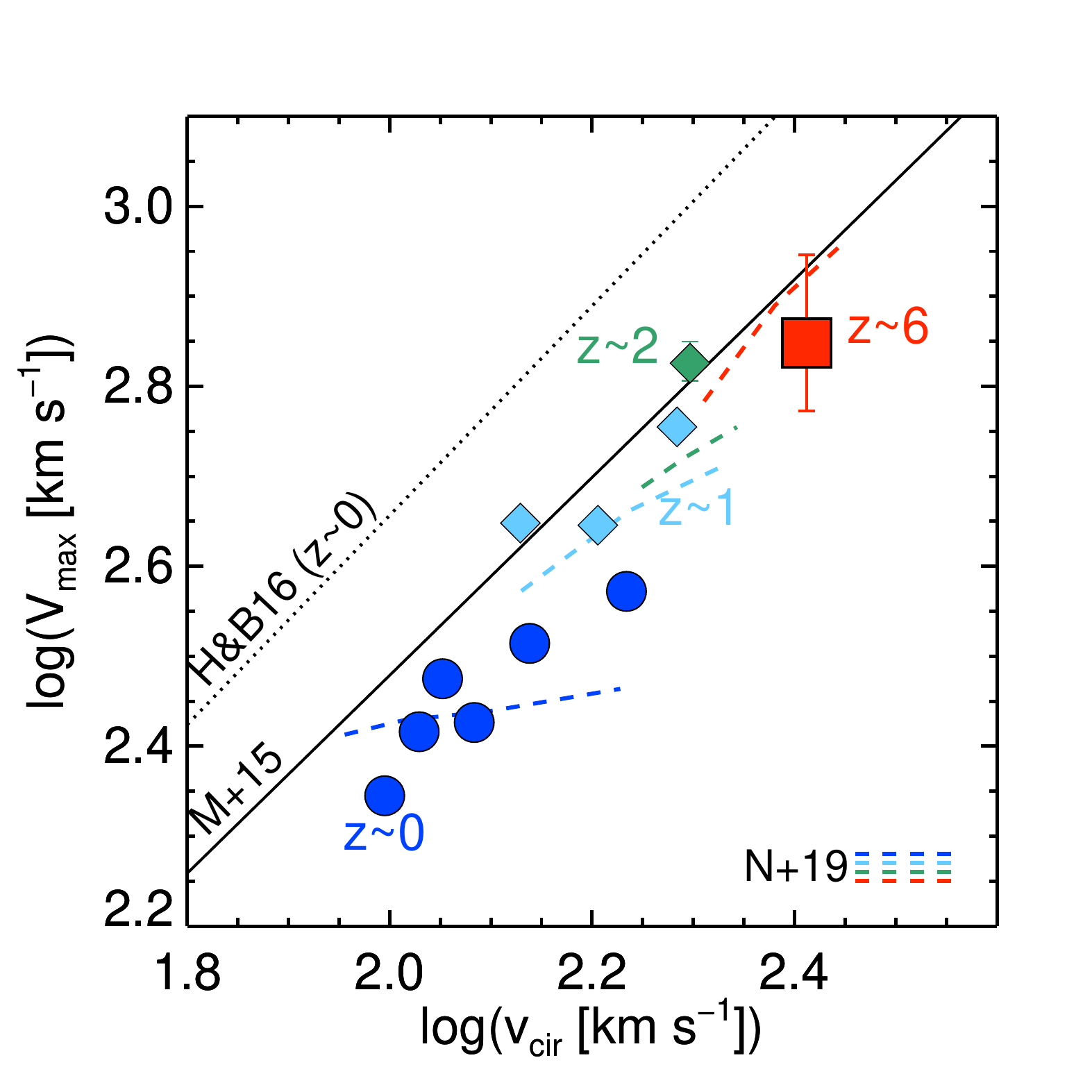}
  \caption{$\vmax$ as a function of the circular velocity \vcir\ that are converted from the stellar mass. The symbols are the same as in Figure \ref{fig:res1}. The solid black line and colored dashed lines represent a theoretical relation at $z=0.5\nn4$ predicted by the FIRE simulations \citep[the flux-weighted average 90th percentile velocity;][]{Muratov:2015} and relations at $z=0$ (blue), $1$ (cyan), $2$ (green), and $6$ (red) predicted by the IllustrisTNG simulation \citep[90th percentile velocity;][]{Nelson.D:2019a}, respectively. The dotted line indicates a relation of extreme-starburst galaxies $z\sim0$ \citet{Heckman:2016}.
  }
  \label{fig:res2}
\end{figure}

Figure \ref{fig:res2} illustrates \vmax\ as a function of \vcir that are calculated from $M_*$ in Section \ref{sec:sample}. Because the galaxies at $z=0\nn6$ have similar $M_*$ and $M_\mathrm{h}$ values (Table \ref{tb:3}), the data points are located in different \vcir\ ranges depending on the redshifts. In the figure, \vmax\ tightly correlates with \vcir\ at $z\sim0$. A correlation with a similar slope at $z\sim0$ is also seen in the cyan diamonds at $z\sim1$. Although only one measurement is available at $z\sim2$ and $z=5\nn6$, respectively, the two data points at $z\sim2\nn6$ appear to follow the relation at $z\sim0\nn1$. Therefore, Figure \ref{fig:res2} suggests a single relation between \vmax\ and \vcir\ that holds over $z\sim0\nn6$. The dotted line indicates a relation at $z=0$ obtained from observations by the Cosmic Origin Spectrograph mounted on the Hubble Space Telescope \citep{Heckman:2016}, which has a similar slope to our measurements. The offset between our data points and the dotted line may arise from the fact that our data points represent the average properties of galaxies at each redshift while their extreme-starburst galaxies have much higher SFR than our galaxies.

\subsection{Redshift evolution of outflow velocities}
\label{sec:results2}

\begin{figure}[t]
  \epsscale{1.2}
  \plotone{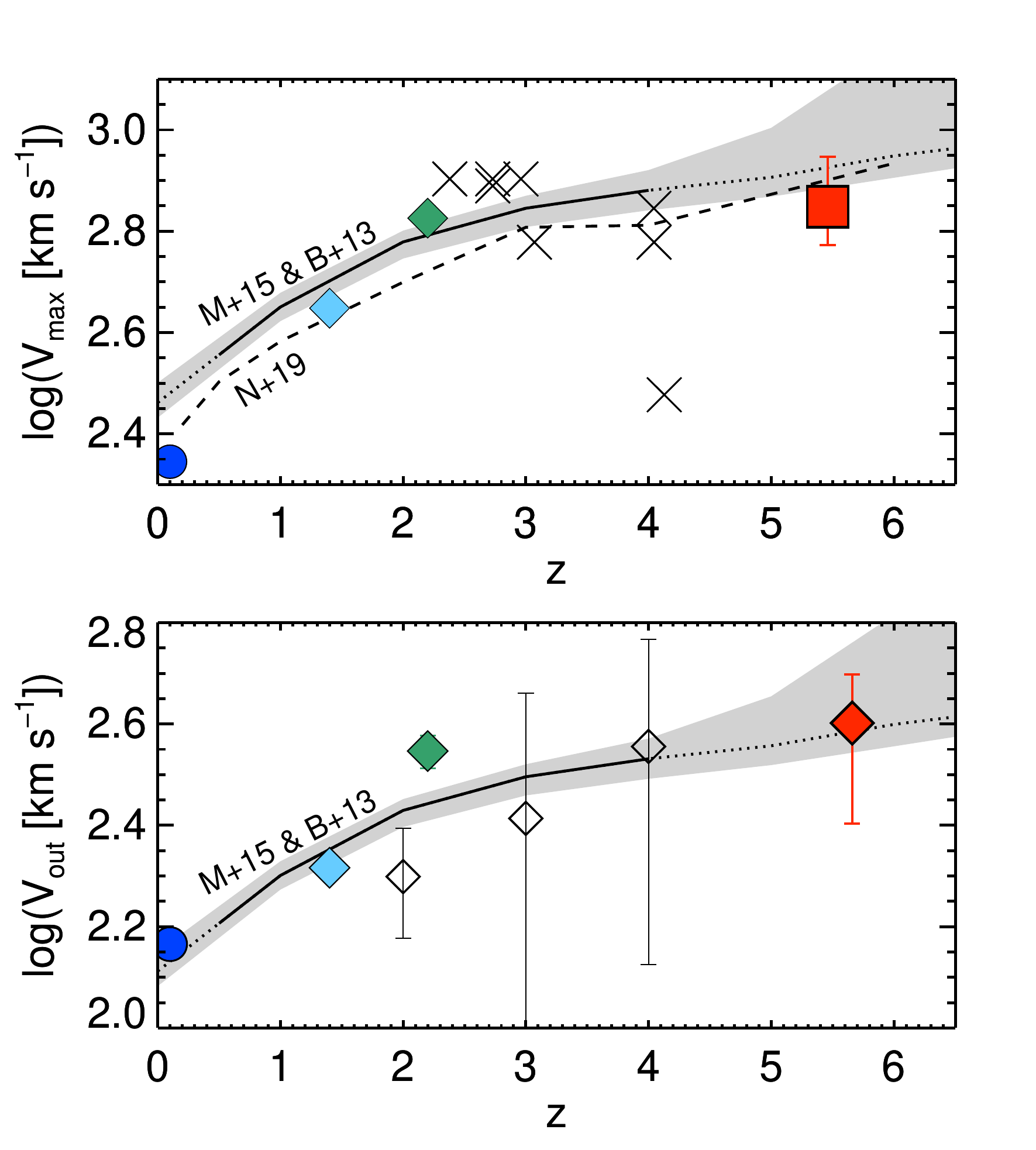}
  \caption{Redshift evolution of \vmax\ (top) and \vout\ (bottom) in the star-forming galaxies with $M_*\sim 10^{10.1}$ \Mo. The colored symbols are the same as in Figure \ref{fig:res1}, but for the red diamond that denotes \vout\ of the galaxies at $z=5\nn6$ measured by a fit of the two-component Gaussian profile to the \CII\ line. To compare the literature, we plot the values of the individual gravitationally-lensed sources in \citet[][ cross]{Jones.T:2013b} and the composite spectra at $z\sim2$, $3$, and $4$ presented by \citet[][ open diamond]{Du.X:2018a}, which have the median stellar masses of $\log (M_*/M_\odot) = 10.00$, $9.87$, and $9.72$, respectively. The solid lines indicate the evolution of the flux-weighted 90th (top) and 50th (bottom) outflow velocities at $M_* = 10^{10.1}$ \Mo\ in the FIRE simulations \citep{Muratov:2015} that we convert from the velocity--\vcir\ relation at $z=0.5\nn4$ using Equations (\ref{eq:0a})--(\ref{eq:0b}) and the SHMR of \citet{Behroozi:2013}. The evolution is extrapolated to $z<0.5$ and $z>4$ (dotted line) and the errors of the SHMR are shown in the shaded regions. The dashed line in the top panel indicates the evolution of the 90th percentile velocity at $M_* = 10^{10.1}$ \Mo\ in the IllustrisTNG simulation \citep{Nelson.D:2019a}.}
  \label{fig:evolution}
\end{figure}

We illustrate the redshift evolution of \vmax\ in star-forming galaxies with $M_*\sim 10^{10.1}$ \Mo\ in the top panel of Figure \ref{fig:evolution}. As shown in Section \ref{sec:results1}, $\vmax$ strongly increases from $z\sim0$ to $2$ and slightly from $z\sim2$ to $6$.
Although many studies measure outflow velocities at fixed redshift, a few studies investigate the redshift evolution of the velocities in wide redshift ranges. \citet{Jones.T:2013b} present the maximum outflow velocity of gravitationally lensed sources at $z\sim2\nn4$ (cross). The stellar masses of these sources are not estimated and the outflow velocity of them is measured in a different manner from ours. However, the sources show similar outflow velocities to our \vmax\ values at $z\sim2$ and $z=5\nn6$, except for a data point of $\vmax \simeq 300$ \kms. \citet{Jones.T:2013b} suggest a decrease in \vmax\ at high redshift that is not statistically significant. We find no decrease at $z=5\nn6$.

\citet{Sugahara.Y:2017a} and \citet{Du.X:2018a} discuss the redshift evolution of the central outflow velocity (\vout) measured with a two-component profile. While the maximum outflow velocity shows the highest velocity that the outflowing gas reaches, the central outflow velocity represents the bulk motion of the outflowing gas. Because the absorption lines include absorption components from outflows and ISM at the systemic velocity, \vout\ should be estimated by a two-component fitting. Although a two-component fitting produces larger errors of the best-fit values than a one-component fitting, we fit a two-component Gaussian profile to the \CII absorption line in the composite spectrum at $z=5\nn6$, in order to compare the \vout\ values with previous studies at $z \lesssim 4$.

The two-component Gaussian profile consists of the systemic and outflow components; \vout\ is defined as the central velocity of the outflow component. This analysis is identical to that used in \citet{Du.X:2018a}. Before the fitting, the composite spectrum is smoothed by a Gaussian kernel so that the spectral resolution become similar to the composite spectrum at $z\gtrsim2$ in \citet{Sugahara.Y:2017a} and \citet{Du.X:2018a}. We also analyze the composite spectrum at $z\sim2$ presented in \citet{Sugahara.Y:2017a}. 

The measured velocities are $\vout=400^{+100}_{-150}$ \kms\ at $z=5\nn6$ and $\vout=352^{+26}_{-27}$ \kms\ at $z\sim2$. In the bottom panel of Figure \ref{fig:evolution}, we plot the measured \vout\ values, showing that the \vout\ redshift evolution has similar features to the \vmax\ evolution: a strong increase from $z\sim0$ to $2$ and no increase from $z\sim2$ to $6$ within the errors. The latter is consistent with a result of \citet{Du.X:2018a}. The \vmax\ and \vout\ values at $z\sim0$, $1$, $2$, and $5\nn6$ are listed in Table \ref{tb:4}.

The open diamonds indicate \vout\ at $z\sim2$, $3$, and $4$ given by \citet{Du.X:2018a}. The \vout\ value at $z=5\nn6$ is comparable to those at $z\sim3$ and $4$ within the marginally large error bars. However, the value at $z\sim2$ denoted by the green diamond is not consistent with the one denoted by the open diamond. In addition, the error bars of the open diamonds are generally larger than those of the filled symbols, in spite of the fact that \citet{Du.X:2018a} stacked a larger number of galaxy spectra than this study and \citet{Sugahara.Y:2017a}. These results may be attributed to the uncertainty of the systemic redshifts in \citet{Du.X:2018a}, who determine the systemic redshifts from the \Lya emission or interstellar absorption lines. When individual spectra are stacked using the systemic redshifts, the uncertainties of the systemic redshifts broaden absorption lines in the composite spectrum. It is possible that this broadened absorption line produces low values and large errors of the best-fit parameters measured with the two-component fitting, which are sensitive to absorption-line profiles. We note that the median stellar masses of the galaxies in \citet{Du.X:2018a} are $\log (M_*/M_\odot) = 10.00$, $9.87$, and $9.72$ at $z\sim2$, $3$, and $4$, respectively, which are less than $M_*$ of our galaxies. It is also possible that this small $M_*$ (i.e., small \vcir) may lead to the low \vout\ value at $z\sim2$.

%\tabletypesize{\scriptsize} %\footnotesize{ %
\begin{deluxetable}{cccccc}
\tablecolumns{5}
\tablewidth{0pt}
\tablecaption{Values of the data points at each redshift in Figure \ref{fig:evolution} \label{tb:4}}
\centering
\tablehead{
\colhead{redshift} \vspace{-0.2cm} & \colhead{\vout} & \colhead{\vmax} & \colhead{$\log(M_*/M_\sun)$\tablenotemark{a}} & \colhead{reference}\\
\colhead{} & \colhead{(\kms)} & \colhead{(\kms)} & \colhead{} & \colhead{}
}
%\decimalcolnumbers
\startdata
$z\sim0$  & $146\pm5.2$ & $221\pm9.9$ & 10.2 & S17 \\
$z\sim1$  & $207\pm5.0$ & $445\pm5.7$ & 10.0 & S17 \\
$z\sim2$  & $352^{+26}_{-27}$\tablenotemark{b} & $673^{+35}_{-33}$\tablenotemark{b} & 10.3 & S17 \\
$z=5\nn6$ & $400^{+100}_{-150}$ & $700^{+180}_{-110}$ & 10.1 & This study \\
\enddata
\tablenotetext{a}{The mean stellar mass of the galaxies.}
\tablenotetext{b}{The outflow velocities at $z\sim2$ are re-measured in this study.}
\tablerefs{S17: \citet{Sugahara.Y:2017a}}
\vspace{-0.7cm}
\end{deluxetable}

\section{Discussion}
\label{sec:discussion}

\subsection{Comparisons with theoretical models}
\label{sec:physical-background}

Recent numerical and zoom-in simulations can be used to predict outflow velocities. These simulations compute energy input to the ISM surrounding SNe and investigate the statistics of galaxy and outflow properties \citep[e.g.,][]{Muratov:2015,Christensen:2016a,Mitchell.P:2018a,Nelson.D:2019a}. Here we compare our results with simulation work that studies the redshift evolution of the outflow velocities. In the figures we convert $M_*$ (\vcir) into \vcir\ ($M_*$) in the simulation work by using Equations (\ref{eq:0a})--(\ref{eq:0b}) and the SHMR in \citet{Behroozi:2013}.

\citet{Nelson.D:2019a} analyze $\sim$20,000 galaxies in the IllustrisTNG simulation to provide statistical relations between outflow and galaxy properties, including SN and AGN feedback. Figure \ref{fig:res2} shows outflow velocities, $v_\mathrm{out,90,r=10\ kpc}$, defined as the 90th percentile of the flux-weighted velocity distribution at a radius of 10 kpc, at $z=0.2$, $1$, $2$, and $6$ in stellar mass ranges similar to observational data points at the redshifts. This theoretical prediction by \citet{Nelson.D:2019a} agrees well with our observational measurements, although the trend at $z\sim0$ is different.

\citet{Muratov:2015} calculate the flux-weighted velocity of the outflowing gas at one quarter of the halo virial radius with the Feedback in Realistic Environments (FIRE) simulations, which computes the thermal and momentum input to the ISM considering the stellar and SN feedback. The outflow velocity in the FIRE simulations tightly correlates with the halo circular velocity and the correlation does not exhibit the significant evolution over $z\sim0.5\nn4$. Figure \ref{fig:res2} illustrates that the correlation in the FIRE simulations at $z=0.5\nn4$ is in good agreement with the tight linear relation which we present in Section \ref{sec:results1}, although the outflow velocity at $z\sim0$ is $\sim0.1$ dex lower than the theoretical prediction.

These agreements with theoretical work support our result that \vmax\ correlates with \vcir\ in a given stellar mass range. However, we note two factors that are important when one compares observations and theories: gas phases and galactocentric radii of outflows. As described in Section \ref{sec:analysis} our observational technique traces low-ionized elements in warm gas ($\lesssim 10^4\ \mathrm{K}$). On the other hand, \citet{Muratov:2015} and \citet{Nelson.D:2019a} compute the \vmax--\vcir($M_*$) relation from outflowing gas with all temperatures. It is noteworthy that a non-negligible fraction of outflowing gas in numerical simulations would be in a hot, diffuse phase which are not observable with optical absorption lines \citep[e.g.,][]{Mitchell.P:2018a,Nelson.D:2019a}. Moreover, in some numerical simulations, outflows in a hot phase tend to exhibit faster velocities than those in cold phases \citep[e.g.,][]{Gallerani.S:2018a,Mitchell.P:2018a}. In addition to gas phases, galactocentric distances of outflowing gas are different between observations and simulations. Observations with the ``down-the-barrel'' technique integrate outflowing-gas absorption along the line of sight and cannot distinguish absorption components at different radii, while most simulations compute outflow velocities at fixed radii. Even among the simulations, \citet{Nelson.D:2019a} and \citet{Muratov:2015} compute velocities at different radii, $10$ \kpc\ and $0.25$ halo virial radius, respectively, despite a radial dependence of outflow velocities \citep{Nelson.D:2019a}. Considering these all factors, it is difficult to interpret the similar \vmax--\vcir\ relation in this work, \citet{Nelson.D:2019a}, and \citet{Muratov:2015}. Nevertheless, the agreement perhaps suggests that multi-phase outflows are accelerated following a common \vmax--\vcir\ relation, irrespective of gas phases.

In Figure \ref{fig:evolution} the black solid and dot-dashed lines indicate the redshift evolution of predicted outflow velocities at $M_* = 10^{10.1}$ \Mo\ based on the results given by the FIRE \citep{Muratov:2015} and the IllustrisTNG \citep{Nelson.D:2019a} simulations, respectively. The evolution based on the simulations is in good agreement with the \vmax\ and \vout\ values in this study and \citet{Sugahara.Y:2017a}, and also with those in \citet{Du.X:2018a} and \citet{Jones.T:2013b}, except for one at $z\sim4$.

These simulations support a monotonic increase in \vmax\ from $z=0$ to $6$ at a fixed stellar mass of $M_*\sim10^{10.1}$ \Mo. This increase will be explained by a monotonic increase in \vcir. While $M_\mathrm{h}$ does not significantly change around $M_*\sim10^{10.1}$ \Mo\ at $z\sim0\nn6$ \citep{Behroozi:2013}, $r_\mathrm{h}$ is proportional to $(1+z)^{-1}$ at a fixed $M_h$ (Equation \ref{eq:0b}). Hence, Equation (\ref{eq:0a}) gives the redshift dependence of the halo circular velocity as $\vcir \propto (1+z)^{0.5}$. Given that \vmax\ has the linear correlation with \vcir\ as shown in Figure \ref{fig:res2}, the redshift evolution in \vmax\ (Figure \ref{fig:evolution}) is explained as reflecting the redshift dependence of \vcir. The power-law index of $0.5$ reproduces the strong increase in \vmax\ from $z\sim0$ to $2$ and the slight increase from $z\sim2$ to $6$.

\subsection{Outflow-velocity correlation with \SFR}
\label{sec:relation-between-sfr}

\begin{figure}[t]
  \epsscale{1.2}
  \plotone{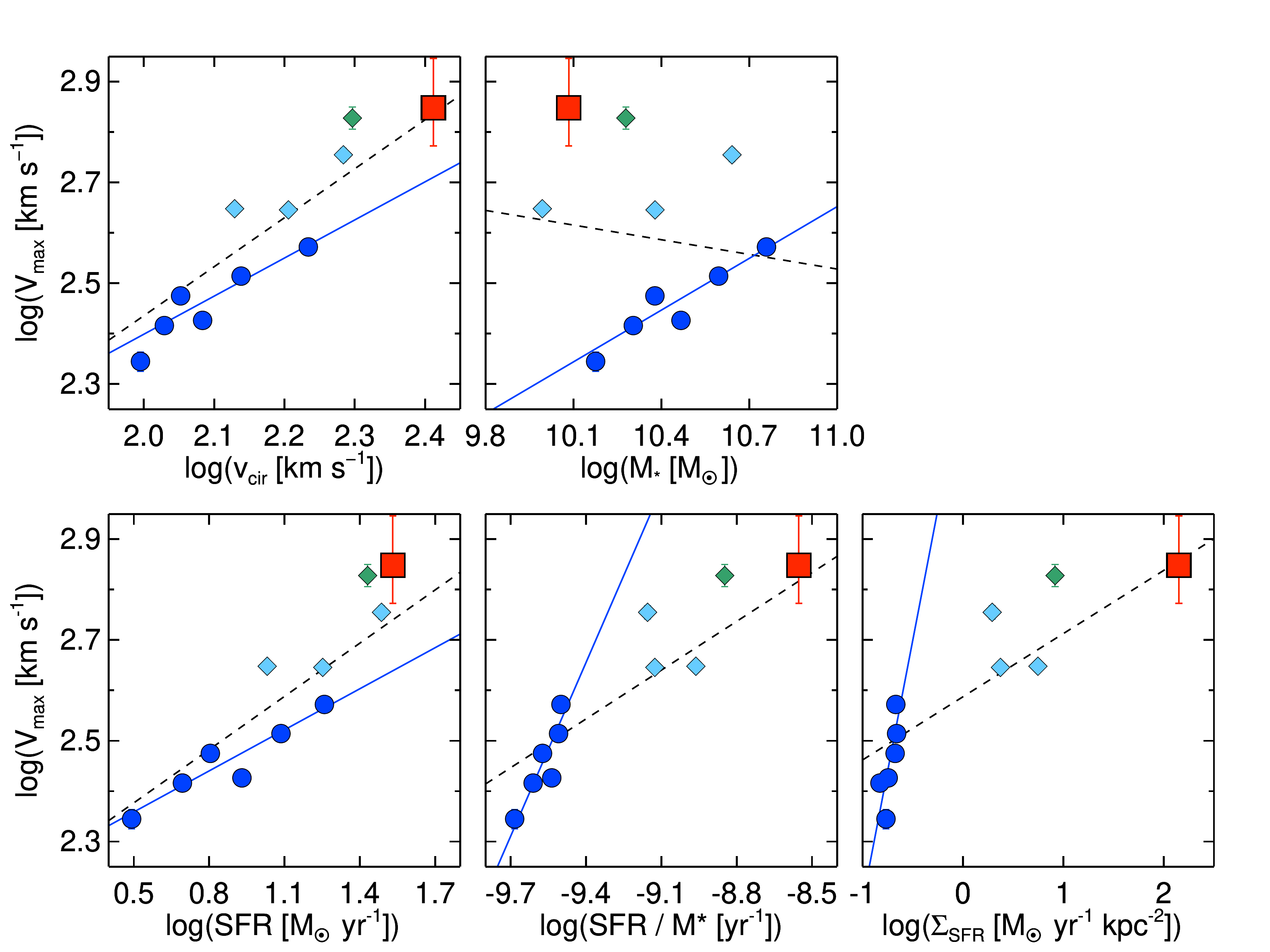}
  \caption{Correlations between \vmax\ and galaxy properties. The symbols are the same as in Figure \ref{fig:res1}. The blue solid lines indicate the best-fit linear relations to the data points at $z\sim0$. The black dashed lines denote the best-fit linear relations to the data points throughout all redshifts.}
  \label{fig:vm-galprop}
\end{figure}

\begin{figure}[t]
  \epsscale{1.2}
  \plotone{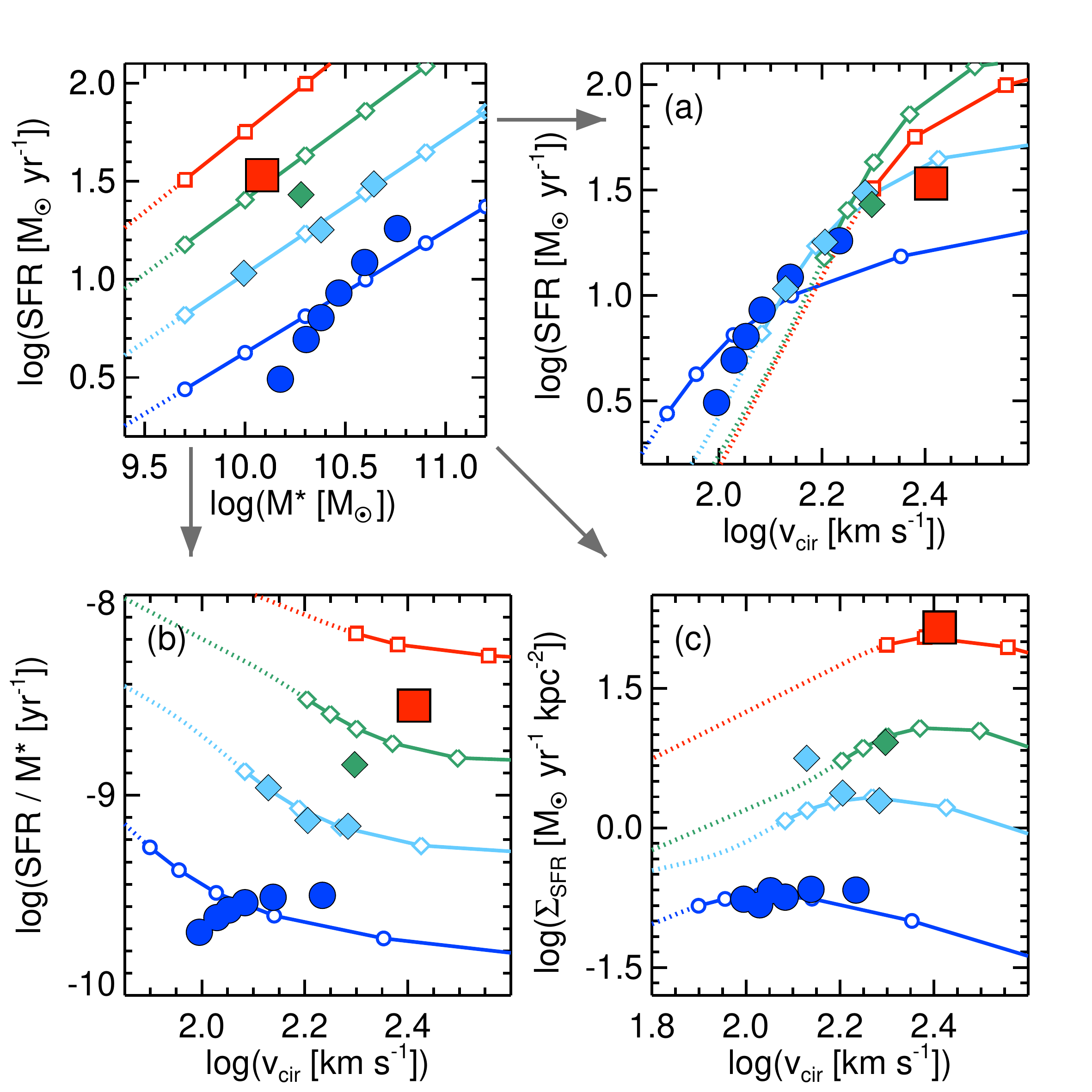}
  \caption{
    Models of the correlations of \SFR, $\SFR/M_*$, and $\Sigma_\SFR$ with \vcir\ for the star-forming main-sequence galaxies.
    The top left panel shows the main sequences at $z\sim0.5$ (blue), $1$ (cyan), $2$ (green), and $6$ (red) that are presented by \citet{Speagle:2014a}. The open symbols on the solid lines are plotted at the intervals of 0.3 dex of $M_*$ for reference. The main sequences are extrapolated to $\log(M_*/\mathrm{M_\odot}) < 9.7$, indicated by the dotted lines. The filled symbols are the same as in Figure \ref{fig:res1}.
    (a) \SFR\ versus \vcir\ where \vcir\ is converted from $M_*$ in the top panel using the SHMR in \citet{Behroozi:2013}. \SFR\ correlates with \vcir\ over $z=0\nn6$.
    (b) $\SFR/M_*$ versus \vcir. 
    (c) $\Sigma_\SFR$ versus \vcir; $\Sigma_\SFR$ is estimated on the assumption that galaxy sizes are proportional to redshifts by $(1+z)^{-1}$ \citep{Shibuya:2015}.
    In the panels (b) and (c), the colored data points, which are in a similar $M_*$ range, exhibit positive correlations. However, the solid lines demonstrate that the relations of the main-sequence galaxies depend on redshifts. Given a \vcir--\vmax\ correlation, the panels (b) and (c) suggest that $\SFR/M_*$ and $\Sigma_\SFR$ are unlikely to be the fundamental parameter.
  }
  \label{fig:sfr-vcir}
\end{figure}

The outflow maximum velocity tightly correlates with the halo circular velocity, but it also has a strong correlation with \SFR. To study the fundamental parameter that determines the outflow velocity over all redshifts, it is worth discussing correlations of \vmax\ with galaxy properties over the wide redshift range. If there exists the fundamental parameter, it should exhibit a single scaling relation with \vmax\ that holds at fixed redshifts and throughout all redshifts.

Figure \ref{fig:vm-galprop} plots \vmax\ as a function of \vcir, $M_*$, \SFR, $\SFR/M_*$, and $\Sigma_\SFR$. Because galaxy sizes are unavailable at high-$z$ due to the spacial resolutions, $\Sigma_{\SFR}$ is estimated on the assumption that galaxy sizes are proportional to redshifts by $(1+z)^{-1}$ \citep{Shibuya:2015}. First, we calculate the Spearman's rank correlations, $r$, between \vmax\ and the galaxy properties over all redshifts. While $M_*$ has no correlation with \vmax, the other galaxy properties exhibit strong correlations of $r=0.81$ (\vcir), $0.78$ (\SFR), $0.90$ ($\SFR/M_*$), and $0.89$ ($\Sigma_\SFR$) with the $> 3\sigma$ significance levels. Next, we perform a linear fitting to the data points at all redshifts and only at $z\sim0$. The best-fit results are illustrated in Figure \ref{fig:vm-galprop}. The best-fit slopes at all redshifts (black line) are positive for \vcir, \SFR, $\SFR/M_*$, and $\Sigma_\SFR$. For \vcir\ and \SFR, the data points show relatively small scatter within $\sim0.1$ dex with respect to the best-fit relation at all redshifts. For $\SFR/M_*$ and $\Sigma_\SFR$, however, the best-fit relations at $z\sim0$ (blue line) have large offsets from the data points at $z\sim1\nn6$, and the slopes of the best-fit relations at $z\sim0$ and at all redshifts are very different from each other. These correlation and linear-fitting tests demonstrate that $M_*$, $\SFR/M_*$ and $\Sigma_\SFR$ show scaling relations at $z\sim0$, but those scaling relations cannot explain the outflow velocity throughout all redshifts. Therefore, \vcir\ and \SFR\ are likely to have the tightest single relations with \vmax\ from $z\sim0$ to $6$.

The strong \vmax-correlations with \vcir\ and \SFR\ imply a strong correlation between \vcir\ and \SFR. To understand the \SFR--\vcir\ relation independent of redshifts, it is helpful to see the distribution of the star-forming main-sequence galaxies on a \SFR--\vcir\ plane. The top left panel of Figure \ref{fig:sfr-vcir} illustrates the main sequences at $z\sim0.5$, $1$, $2$, and $6$ that are presented by \citet{Speagle:2014a}. The galaxies in this work and \citet{Sugahara.Y:2017a} have similar stellar masses in the range of $10.0 < \log(M_*/\mathrm{M_\odot}) < 11.0$. We note that the main sequence at $z\sim0.5$ (blue line) agrees with galaxies at $z\sim0.1$ (blue circles) because \citet{Sugahara.Y:2017a} construct the $z\sim0$ sample with the highly star-forming galaxies to analyze the \NaID absorption line. By converting $M_*$ into \vcir\ with the method in Section \ref{sec:sample}, we plot the main sequences on a \SFR--\vcir\ plane in the panel (a) of Figure \ref{fig:sfr-vcir}. They show similar positive relations at all redshifts, leading to a positive correlation of the main-sequence galaxies, irrespective of redshifts. The data points indeed exhibit a strong positive correlation ($r=0.99$) at the $5.8\sigma$ significance level. This result naturally explains that \vmax\ has a correlation with \vcir\ and \SFR\ simultaneously. In other words, constraining the fundamental parameter requires more measurements in a wide range of the stellar masses, \SFR, and redshifts.

In the panels (b) and (c) of Figure \ref{fig:sfr-vcir}, we plot the main sequences on $\SFR/M_*$--\vcir\ and $\Sigma_\SFR$--\vcir\ planes, respectively, which are useful to understand Figure \ref{fig:vm-galprop}. Contrary to those on the \SFR--\vcir\ plane, the main sequences have offsets in the positive direction from low to high redshifts. This demonstrates that the apparent positive correlation of the data points on the $\SFR/M_*$--\vcir\ and $\Sigma_\SFR$--\vcir\ planes are simply because the galaxies have the similar stellar masses. Given a redshift-independent correlation between \vcir\ and \vmax\ as discussed above, Figure \ref{fig:sfr-vcir} illustrates scaling relations between \vmax\ and galaxy properties at each redshift, which clearly reproduces the distribution of the data points in Figure \ref{fig:vm-galprop}. These simple models suggest that \vmax--$\SFR/M_*$ and \vmax--$\Sigma_\SFR$ relations of the main-sequence galaxies depend on redshifts, namely, that $\SFR/M_*$ and $\Sigma_\SFR$ are unlikely to be the fundamental parameter.

 The parameters which most strongly correlate with \vmax\ are \vcir\ and \SFR, suggesting that the fundamental parameter is \vcir\ or \SFR. This result agrees with previous observational studies that present positive correlations of \vmax\ with $M_*$ \citep{Martin:2005, Rubin:2014, Erb:2012} or \SFR\ \citep{Kornei:2012, Heckman:2015, Heckman:2016}. In many cases, the outflow properties are assumed to be connected with star-forming activities in galaxies. However, \vcir\ affects \SFR\ through the halo accretion rate \citep[e.g.,][]{Harikane.Y:2018a,Tacchella.S:2018a} and this process contributes to form the \SFR--\vcir\ correlation in Figure \ref{fig:sfr-vcir}. Thus, because \vcir\ represents two key parameters for the outflow velocity, the gravitational potential and the star-forming activity, it is important to consider the possibility that \vcir\ is the fundamental parameter to determine the outflow velocity.

\subsection{Lyman-continuum leakage}
\label{sec:lyman-cont-leak}

\begin{figure}[t]
  \epsscale{1.2}
  \plotone{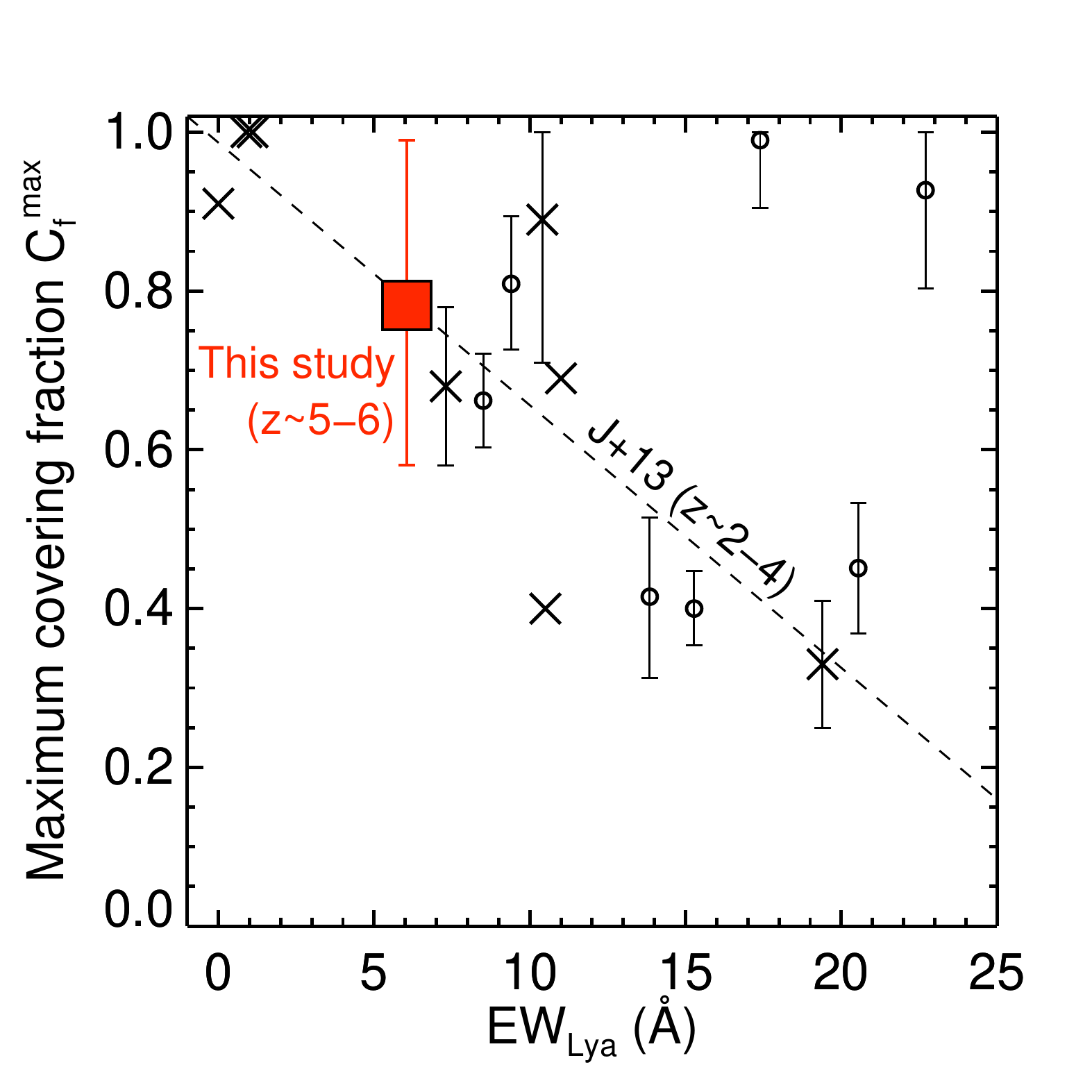}
  \caption{The maximum covering fraction $\Cfmax$ as a function of the \Lya equivalent width. The red square denotes the result at $z=5\nn6$. The crosses and the open circles indicate the values of gravitationally-lensed sources at $z\sim2\nn4$ \citep{Jones.T:2013b} and $z\sim4\nn5$ \citep{Leethochawalit.N:2016a}, respectively. The dashed line is the best-fit linear relation to the crosses \citep{Jones.T:2013b}.}
  \label{fig:fabs}
\end{figure}

The redshift $z=5\nn6$ is near the end of the cosmic reionization, when the neutral IGM has been ionized. Plausible ionizing sources are young, low-mass galaxies \citep[e.g.,][]{Robertson:2015,Ishigaki.M:2018b}, but their contribution is still a matter of debate. A key physical parameter is the escape fraction of the Lyman-continuum (LyC) photons from galaxies (\fescc). Because it is possible that the outflows help increase \fescc\ by creating holes in the neutral ISM from which the LyC photons can escape, the high-$z$ galaxies with outflows in this work are appropriate for the reionization study. However, direct measurements of \fescc\ are challenging for galaxies at $z=5\nn6$ because the LyC photons almost disappear by ionizing the neutral IGM. In this section, we discuss the \fescc\ value of our galaxies at $z=5\nn6$ with two indirect methods regarding the absorption lines.

In the first method, we calculate the covering fraction from the metal absorption lines. In cases of the optically-thick outflowing gas, absorption lines are saturated and the line depth gives the covering fractions. Assuming that the low-ionized elements are associated with the neutral-hydrogen gas, \citet{Jones.T:2013b} evaluate the maximum covering fraction (\Cfmax) of the low-ionized elements from the low-ionized absorption lines as an upper limit of \fescc. Because our composite spectrum has the low continuum S/N, we define \Cfmax\ as $\Cfmax = 1-F_\mathrm{SiII}$, where $F_\mathrm{SiII}$ is the median flux density of the \SiII line from $-350$ to $-100$ \kms\ in the normalized spectrum. Its error is calculated with the parametric bootstrap method based on the spectral noise. The measured value is $\Cfmax = 0.8\pm0.2$. We note that this \Cfmax\ value is likely smaller than the value evaluated by the method in \citet{Jones.T:2013b} because our \Cfmax\ value is calculated in the wide velocity range of 250 \kms. We additionally measure the \Lya equivalent width (\EWLya) of the composite spectrum to be $\EWLya=6.05\pm0.45$ \AA, using the emission strength from the stellar continuum at 1216--1221 \AA.

Figure \ref{fig:fabs} illustrates \Cfmax\ as a function of \EWLya. Our measurement at $z=5\nn6$ (red square) is consistent with previous results \citep{Jones.T:2013b, Leethochawalit.N:2016a} and on the linear relation at $z\sim2\nn4$ presented by \citet[][dashed line]{Jones.T:2013b}. This is the first observational result showing that the linear relation between \Cfmax\ and \EWLya\ holds even at $z>5$, provided that the relation is independent of the stellar mass. Using the \Cfmax\ value corresponding to $\EWLya=6.05$ \AA\ on the relation, we obtain an upper limit of \fescc\ to be $\simeq 0.2$. This secure upper limit is too weak to constrain models where bright galaxies contribute to the cosmic reionization \citep[e.g., $\sim10\%$;][]{Sharma.M:2017a}. However, \citet{Jones.T:2013b} emphasize that the property derived by this method is an upper limit. Following an equation derived by \citet{Chisholm.J:2018b}, who propose indirect estimations of \fescc\ using local LyC leaking galaxies, we obtain $\fescc \lesssim 0.5-0.6\Cfmax = 0.02$. Hence, the intrinsic \fescc\ is likely much lower than the upper-limit value.

In the second method, we consider the shape of the absorption-line profile using the outflow velocities. \citet{Chisholm.J:2017a} calculate the ratio of the maximum outflow velocity to the central outflow velocity ($v_{90}/v_\mathrm{cen}$) of galaxies at $z=0$. They find that the LyC leaking galaxies exhibit smaller ratios, $v_{90}/v_\mathrm{cen}\lesssim 5$, than galaxies without LyC leakage, although there are several galaxies with $v_{90}/v_\mathrm{cen} < 5$ but $\fescc = 0$. Here we use $|\vmax/v_0|$ for an alternative to $v_{90}/v_\mathrm{cen}$ used in \citet{Chisholm.J:2017a}. The ratio for the galaxies at $z=5\nn6$ is obtained to be $|\vmax/v_0| = 2.0\pm0.2$. This result suggests that the galaxies at $z=5\nn6$ are the LyC leaking galaxies, in contrast to the result of the first method. Further studies on both the LyC photons and the absorption-line properties will provide key quantities to address the challenge of estimating \fescc\ for galaxies at the epoch of reionization.

\section{Conclusion}
\label{sec:conclusion}
We study the outflow velocities of star-forming galaxies at $z=5\nn6$ and discuss the redshift evolution of the outflow velocities from $z\sim0$ to $6$ by analyzing rest-frame FUV spectra of seven LBGs at $z=5\nn6$ taken by DEIMOS available to date. We construct a high-S/N composite FUV spectrum based on the systemic redshifts determined by ALMA [\CIIx] 158 $\mu$m observations \citep{Capak:2015} to fit a line profile to the \SiII$\lambda1260$, \CII$\lambda1335$, and \SiIV$\lambda\lambda1394,1403$ absorption lines. One of the best-fit parameters $v_0$, the central velocity of the line profile, is significantly negative, which implies that the absorption lines are blueshifted by the outflows.

The maximum outflow velocity \vmax\ is measured from the best-fit parameters. The \vmax\ values for the low-ionized lines (\SiII and \CII) are comparable to the one for the high-ionized line (\SiIV), within the moderately large errors. By a simultaneous fit to the \SiII\ and \CII\ lines, we obtain $\vmax = 700^{+180}_{-110}$ \kms, which is higher than those at $z\sim0$ and comparable to the one at $z\sim2$ presented by \citet{Sugahara.Y:2017a}. This result represents the redshift evolution of \vmax\ that strongly increases from $z\sim0$ to $2$ and weakly increases from $z\sim2$ to $6$, at the fixed stellar mass of $\log (M_*/\mathrm{M_\sun}) \sim10.1$. We additionally measure the central outflow velocity (\vout) by fitting a two-component Gaussian profile to the \CII line, and confirm that the redshift evolution of \vout\ is similar to the \vmax\ evolution.

Over $z\sim0\nn6$, $\log \vmax$ is linearly correlated with the halo circular velocity ($\log \vcir$) that are estimated from the stellar mass. This linear correlation can explain the increasing features of the \vmax\ evolution because \vcir\ is proportional to $(1+z)^{0.5}$ for the galaxies with $\log (M_*/\mathrm{M_\sun}) \sim10.1$, at which the halo mass is almost constant over $z\sim0\nn6$ \citep{Behroozi:2013}. In addition, the correlation between \vmax\ and \vcir\ is in good agreement with a relation predicted by the IllustrisTNG \citep{Nelson.D:2019a} and the FIRE \citep{Muratov:2015} simulations. Although there are differences of gas phases and galactocentric radii between the simulation and observational work, this good agreement perhaps suggest that the multi-phase outflows are driven by a common \vmax--\vcir\ relation.

The outflow maximum velocity \vmax\ strongly correlates with \vcir, \SFR, $\SFR/M_*$, and $\Sigma_\SFR$ over $z=0\nn6$. Moreover, on the \vmax--\vcir\ and \vmax--\SFR\ planes, the linear scaling relations at $z=0$ explain the whole distribution from $z=0$ to $6$. Given that the \vmax--\vcir\ relation holds at any redshifts, the models of the star-forming main sequences reproduce the relation between \vmax\ and galaxy properties at $z=0\nn6$. For these reasons, \vcir\ or \SFR\ are likely to be the fundamental parameter to determine \vmax\ with a single relation throughout all redshifts. Considering that \vcir\ has an impact on \SFR\ through the halo accretion rate, it is possible that \vcir\ is the fundamental parameter.

Absorption-line profiles are also used for indirect estimations of the escape fraction of the LyC photons (\fescc). We find that the maximum covering fraction of the \SiII line and the \Lya equivalent width of the composite spectrum at $z=5\nn6$ are consistent with a relation at $z\sim2\nn4$. The intrinsic \fescc\ would be much lower than the secure upper limit $\fescc < 0.2$, while the ratio $|\vmax/v_0|$ is comparable to the values of the local LyC leaking galaxies.

\acknowledgements

We thank Xinnan Du, Kate Rubin, John Chisholm, L\'{e}o Michel-Dansac, Timothy M. Heckman, and Hidenobu Yajima for very useful discussion and comments. We wish to thank the referee for constructive and valuable suggestions for improvement.
We acknowledge Peter Capak, the PI of the data in this work. The data presented herein were obtained at the W. M. Keck Observatory, which is operated as a scientific partnership among the California Institute of Technology, the University of California and the National Aeronautics and Space Administration. The Observatory was made possible by the generous financial support of the W. M. Keck Foundation.
This research has made use of the Keck Observatory Archive (KOA), which is operated by the W. M. Keck Observatory and the NASA Exoplanet Science Institute (NExScI), under contract with the National Aeronautics and Space Administration.
The authors wish to recognize and acknowledge the very significant cultural role and reverence that the summit of Maunakea has always had within the indigenous Hawaiian community.  We are most fortunate to have the opportunity to conduct observations from this mountain.
This work is supported by World Premier International Research Center Initiative (WPI Initiative), MEXT, Japan, and KAKENHI (15H02064, 17H01110, and 17H01114) Grant-in-Aid for Scientific Research (A) through Japan Society for the Promotion of Science. Y.S. acknowledges support from the JSPS through the JSPS Research Fellowship for Young Scientists.

\bibliographystyle{aasjournal}
\bibliography{$HOME/Documents/set_TeX/reference} %$

\end{document}